\documentclass[useAMS,usenatbib,usegraphicx]{mn2e}

\usepackage{fixltx2e}

\def\'#1{\ifx#1i{\accent"13\i}\else{\accent"13#1}\fi}

\def\alamenos#1{$^{-#1}$}
\def\ala#1{$^{#1}$}

\def\diezala#1{10$^{#1}$}

\def\aap{A\& A}

\def\apj{ApJ}
\def\apjs{ApJS}
\def\apjl{ApJL}
\def\apss{Ap\& SS}

\def\araa{ARA\&A}

\def\mnras{{MNRAS}}

\def\'#1{\ifx#1i{\accent"13\i}\else{\accent"13#1}\fi}
\def\alamenos#1{$^{-#1}$}
\def\ala#1{$^{#1}$}

\def\diezala#1{10$^{#1}$}

\def\VS{{V\'azquez-Semadeni}}

\def\Msun{$M_\odot$}

\def\BP{Ballesteros-Paredes}

\title[Gravity or Turbulence?] {Gravity or Turbulence? The velocity
  dispersion-size relation}

 \author[Ballesteros-Paredes et al. ]
{ \parbox{7.0in}{
Javier Ballesteros-Paredes\ala 1\thanks{e-mail: {\tt j.ballesteros@crya.unam.mx}},
Lee W. Hartmann\ala 2,
Enrique V\'azquez-Semadeni\ala 1,
Fabian Heitsch\ala 3, and
Manuel A. Zamora-Avil\'es\ala 1 \\
}\\
\ala 1 Centro de Radioastronom\'ia y Astrof\'isica,
            Universidad Nacional Aut\'onoma de M\'exico, \\
            Apdo. Postal 72-3 (Xangari), Morelia,
            Michoc\'an 58089, M\'exico \\
     \ala 2 Department of Astronomy, University of Michigan,  500
           Church Street, Ann Arbor, MI 48105, USA \\
     \ala 3 Department of Physics and Astronomy, University of North
            Carolina Chapel Hill, \\ CB 3255, Phillips Hall, Chapel
            Hill, NC 27599, USA \\
}

\begin{document}


\date{Accepted by MNRAS, \today}

\pagerange{\pageref{firstpage}--\pageref{lastpage}} \pubyear{2010}

\maketitle

\label{firstpage}

\begin{abstract}

We discuss the nature of the velocity dispersion vs. size relation for
molecular clouds.  In particular, we add to previous observational
results showing that the velocity dispersions in molecular clouds and
cores are not purely functions of spatial scale but involve surface
gas densities as well.  We emphasize that hydrodynamic turbulence is
required to produce the first condensations in the progenitor medium.
However, as the cloud is forming, it also becomes bound, and
gravitational accelerations dominate the motions. { Energy
conservation in this case implies $|E_g| \sim E_k$, in agreement with
observational data, and providing an interpretation for two recent
observational results: } the scatter in the $\delta v-R$ plane, and
the dependence of the velocity dispersion on the surface density
${\delta v^2/ R} \propto \Sigma$.  We argue that the observational
data are consistent with molecular clouds in a state of {\it
hierarchical gravitational collapse}, i.e., developing local centers
of collapse throughout the whole cloud while the cloud itself is
collapsing, { and making equilibrium unnecessary at all stages
prior to the formation of actual stars.  Finally, we discuss how this
mechanism need not be in conflict with the observed star formation
rate.}

\end{abstract}


\section{Introduction}\label{sec:intro}

Almost 30 years ago, \citet{Larson81} suggested that two scaling
relations exist for molecular clouds, one for the velocity dispersion
$\delta v$ and the other for the mean density ${\rho}$. These
have the form

\begin{equation}
  \rho \propto R^\alpha,
  \label{eq:larson:rho-r} 
\end{equation}
\begin{equation}
  \delta v \propto R^\beta,
  \label{eq:larson:dv-r} 
\end{equation}
where $R$ is the size of the cloud.  The most commonly accepted values
of the exponents are\footnote{Note that the original values for the
  exponents reported by \citet{Larson81} are $\alpha \sim -1.1$ and
  $\beta \sim 0.38$.}  $\alpha \sim -1$ and $\beta \sim 0.5$
\citep[see][ and references therein]{Solomon_etal87, Blitz93,
  BP_etal07, {McKee_Ostriker07}}.  The first relation implies that the
mean column density of the gas, $\Sigma = \rho R$, is roughly constant
for the whole ensemble of clouds. This result has been challenged by
various authors; for instance, \citet{Kegel89} pointed out that it may
be the result of various selection effects.  \citet{Scalo90} showed
that the study by \citet{Solomon_etal87} was sensitive only to a
limited dynamical range of column densities.  Moreover, \cite{VBR97}
used numerical simulations of a 1-kpc$^2$ piece of the Galaxy to argue
that the mean density-size relation does not hold when no
detectability limitations exist. \citet{BM02} used three-dimensional
simulations to confirm that clouds can have different mean densities,
and that the density-size relation appears when clouds are observed
with particular tracers, since they must have at least a minimum
column density in order to be detectable.

On the other hand, the velocity-size relation (equation
\ref{eq:larson:dv-r}) has often been assumed to be real.  One reason
for this is probably that the relation is similar to what that which
might be expected from studies of fluid turbulence: the expected value
for incompressible turbulence is $\beta \sim 1/3$, while a turbulent
fluid dominated by shocks might exhibit $\beta \sim 1/2$ \citep[e.g.,
][{ and references therein}]{Elmegreen_Scalo04, McKee_Ostriker07}.
As the interstellar medium is highly compressible, and subject to
strong shocks of stellar winds, supernovae, spiral arms, etc., one
might then expect that the velocity dispersion-size scaling relation
with $\beta = 1/2$ should be valid.  And of course, supersonic
turbulent simulations, as well as analytical calculations of molecular
clouds performed over the last decade, have supported the idea of a
velocity dispersion-size relation with $\beta \sim 1/2$ \citep[e.g.,
][ etc]{VBR97, Padoan_Nordlund99, BVS99, BM02, Padoan_Nordlund02,
  Krumholz_McKee05, Field_etal08}.

However, if Larson's original dataset, as well as that of
\cite{Solomon_etal87} were limited in the range of column densities
observed, why should only one of these correlations be affected but
not the other one?  In principle, one may expect that the velocity
dispersion-size relation is also implicitly a result for a limited
range of column densities.

Recently, \citet{Heyer_etal09} observed an ensemble of molecular
clouds with the 14~m FCRAO telescope.  Because these data have
improved sensitivity and better spectral and angular resolution, they
reach much larger surface densities than the data in \citet{Larson81}
and \citet{Solomon_etal87}.  \citet{Heyer_etal09} found that the
velocity dispersion does not depend simply on size scale, but on the
square root of the column density as well.  They went on to point out
that the revised relation,

\begin{equation}
\delta v \propto \Sigma^{1/2} R^{1/2},
\label{eq:quasivirial}
\end{equation}
was that to be expected for clouds in gravitational {\it equilibrium}.
In addition, while the derived masses for the clouds were a factor of
a few lower than expected for Virial equilibrium, \citet{Heyer_etal09}
stated that uncertainties in their mass estimates still allowed for
consistency with equilibrium states.  \citet{Larson81} actually came
to a similar conclusion even with his limited dataset; he argued that
the clouds are mostly gravitationally bound and in approximate Virial
equilibrium.  However, { since the clouds' column densities vary by
  over three orders of magnitude (from $\sim$~$10^{21}$~cm\alamenos
  2 in the most diffuse clouds, to $\sim$~\diezala{24}~cm\alamenos2
  for infrared dark clouds)}, then the simple form of the
velocity-size relation, equation (\ref{eq:larson:dv-r}), may not be
the most appropriate form for molecular clouds.

To further explore the correlation between the velocity dispersion and
size for molecular cloud cores, in \S\ref{sec:obs} we compile recent
data from the literature in order to show that a unique trend for the
scaling of the velocity dispersion { with size} does not appear to
exist, specifically when we include recent sensitive observations of
massive cores.  Instead, relation (\ref{eq:quasivirial}) seems to hold
in all cases.

The question is then, what is the origin of the velocity-surface
density-radius correlation?  { In the past, it has been proposed
  that this relationship arises from a condition of hydrostatic
  equilibrium applying to the clouds and dense clumps \citep{Elm89,
    MT03, Field_etal10}. In this case, it is assumed that the clouds
  are confined by an external bounding pressure, {\it which is
    estimated from the clouds' column density,} and that the role of
  the turbulent motions within the clouds is to provide support. }
However, as summarized by \citet{BVS99} and \citet{BP06}, the complex
non-linear, large-scale, and anisotropic nature of turbulent motions
implies that they do not necessarily provide support, but rather cause
continuous morphing and reshaping of molecular clouds, and contribute,
or perhaps are even driven by, the clouds' gravitational collapse
\citep{VS_etal08}. In particular, it is difficult to see how such an
irregular and locally anisotropic velocity field could ``know'' how to
adjust the magnitude and orientation of the turbulent motions to
maintain clouds in approximate equilibrium for several free-fall
times.  Indeed, it is difficult to argue that global equilibrium is
maintained, given evidence for age spreads in stellar populations that
are smaller than the lateral crossing times of the clouds
\citep{BHV99, HBB01}.  As we discuss in the present contribution, the
observed correlations can be explained as long as the velocity
dispersions result from gravitational acceleration, without requiring
equilibrium { at any stage prior to the formation of an actual
  star.}

\section{Velocity dispersion vs size relation?}\label{sec:obs}

Although it is frequently argued that molecular clouds and their cores
usually exhibit a relation like eq. (\ref{eq:larson:dv-r}), after the
study by \citet{Caselli_Myers95}, and probably more clearly that by
\citet{Plume_etal97}, it became somehow recognized that massive cores
may exhibit a shallower slope than the frequently quoted $\delta
v\propto r^{1/2}$.  Moreover, the various available datasets have not
been plotted all together.

\begin{figure*}
\includegraphics[width=0.55\hsize]{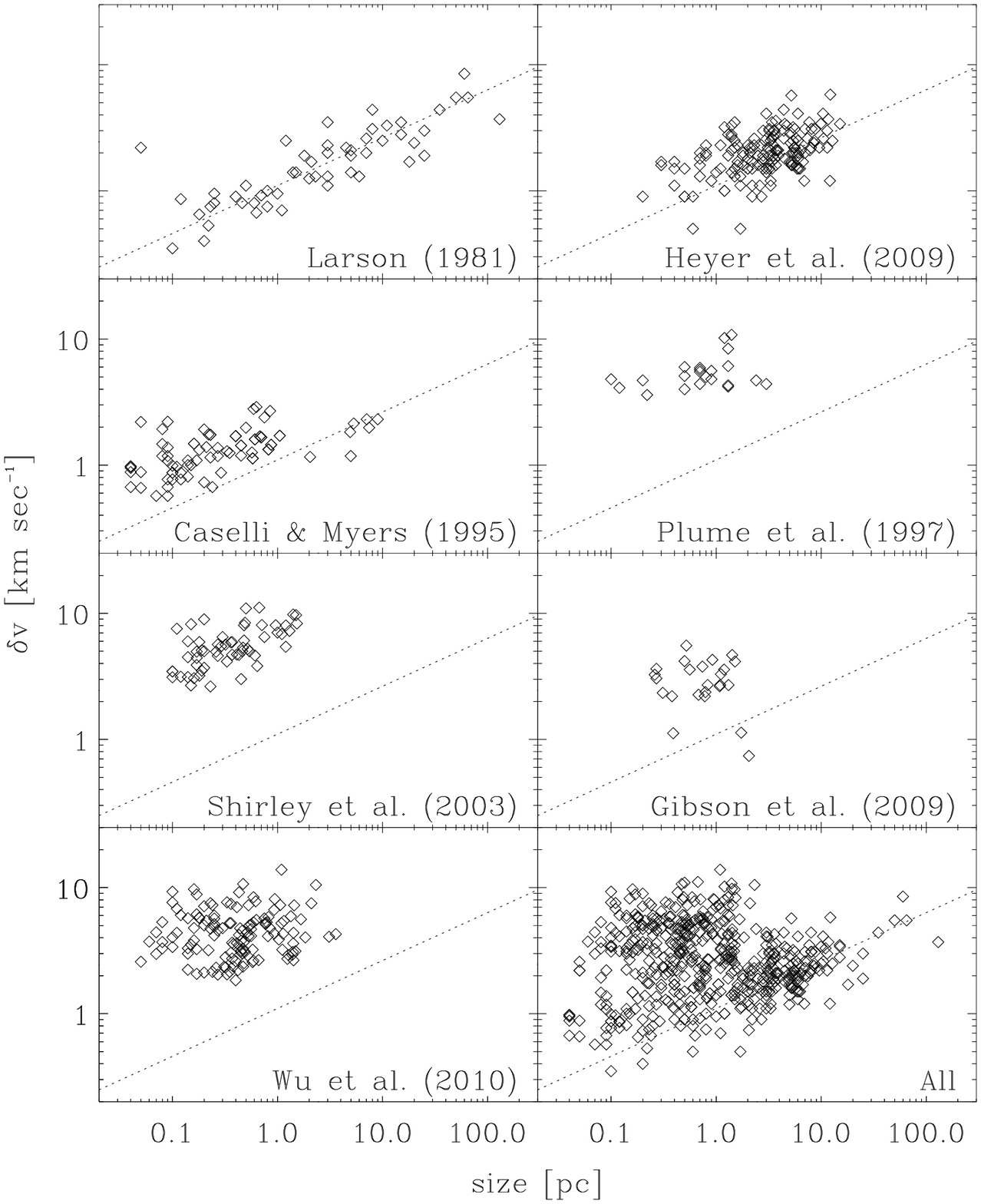}
\caption{\label{fig:dv_r_obs} Velocity dispersion-size space for
  observed cores.  Data compiled from \citet{Larson81,
  Caselli_Myers95, Plume_etal97, Shirley_etal03, Gibson_etal09,
  Heyer_etal09} and \citet{Wu_etal10}. Note that, while large CO
  clouds from the survey by \citet{Heyer_etal09} exhibit the typical
  Larson relationship, dense massive cores are located above this
  relationship.  All together (lower-right panel), the data do not
  exhibit a relation between the velocity dispersion and size.}
\end{figure*}

In Fig.~\ref{fig:dv_r_obs} we plot the velocity dispersions as a
function of size for the dense cores given by \citet{Caselli_Myers95,
  Plume_etal97, Shirley_etal03, Gibson_etal09} and \citet{Wu_etal10}.
For comparison, we have included also the data points of the original
work of \citet{Larson81}, as well as the recent data by
\citet{Heyer_etal09}, neither of which focused particularly on dense
massive cores.  The dotted lines in this figure represent Larson's fit
to his data, i.e.,

\begin{equation}
\biggl({\delta v\over {\rm km\ s^{-1}}}\biggr) = 1.1 \biggl({L\over
    {\rm pc}}\biggr)^{0.38}.
\label{eq:larsonfit}
\end{equation}

We observe that, while in general terms the typical CO clouds observed
by \citet{Heyer_etal09} lie close to Larson's relation, this is
clearly not the case for the dense, massive cores, which exhibit large
velocity dispersions for their relatively small sizes.  Although the
deviation is only marginal for the Orion cores observed by
\citet{Caselli_Myers95}, it is clearer for the more massive compact
cores cores reported by the rest of the datasets.

\begin{figure}
\includegraphics[width=1\hsize]{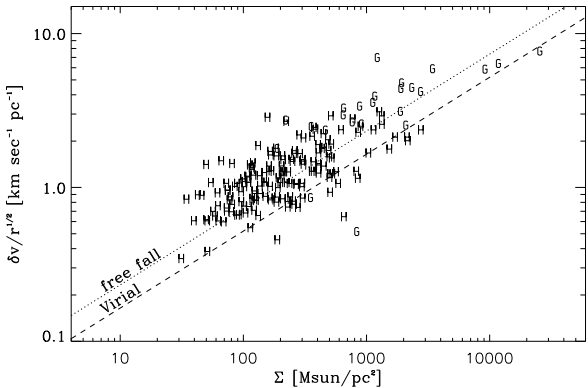}
\caption{\label{fig:heyer_relation} Heyer's relation  ($\delta
  v/r^{1/2}$ vs. surface density $\Sigma$) for the clouds reported in
  \citet{Heyer_etal09, Gibson_etal09}. Note that the massive
  cores follow the tendency showed by \citet{Heyer_etal09}, but they
  are at the large column density end of that relationship, ranging
  from 100 to $\sim 25,000$~\Msun pc\alamenos 2.}
\end{figure}

In Fig.~\ref{fig:heyer_relation} we plot the ``Heyer relation'',
$\delta v/r^{1/2}$ vs. surface density $\Sigma$.  Unfortunately, the
observations of massive dense cores, particularly from infrared dark
clouds, are very recent, and only few cores have independent mass
estimations.  From the list of references given above, only that by
\citet{Larson81, Heyer_etal09} and \citet{Gibson_etal09} list the
masses of their observed cores independent of the Virial mass, and
thus, only these data can be plotted.  In this figure, the data points
from the samples by \citet{Heyer_etal09} and \citet{Gibson_etal09} are
denoted by H and G, respectively.  The long-dashed and dotted lines
respectively represent the loci of structures in Virial equilibrium
and structures undergoing free-fall. From this figure, it is clear
that the massive cores from \citet{Gibson_etal09}  span over two
orders of magnitude in column density, from $\sim$ 100 to a few times
\diezala 4~\Msun\ pc\alamenos 2, the latter being a factor of 10
larger than the maximum column density in the \citet{Heyer_etal09}
data.  We note that 
{ The entire dataset is seen to be reasonably well fitted by a
relation of the form proposed by Heyer and in fact, at face value,
seems to agree better with the free-fall regime than with Virial
equilibrium. }

{ It is important to note that the free-fall relation actually
implies {\it larger} velocity dispersions than the Virial equilibrium
one, contrary to the very common interpretation that velocities higher
than Virial imply that the clouds are unbound. From this discussion,
we see that they can mean chaotic infalling motions instead.}

\section{Discussion}

\subsection{The $\delta v-R$ relationship, a consequence of the 
   role of gravity in the formation and evolution of molecular
   clouds}\label{sec:discussion}

In order to understand the nature of the $\delta v-R$ relation it is
important to understand how molecular clouds are formed and how they
evolve.  

In the last few years, a number of authors have supported the idea
that molecular clouds are formed out of atomic gas when large-scale
streams collide at transonic speeds\footnote{These ideas have been
  developed specifically for understanding star formation in the solar
  neighborhood, where most of the gas is atomic.  In other regions
  such as the molecular ring, large-scale flows in molecular gas can
  also produce dense star-forming regions.}  \citep[e.g., ][]{BVS99,
  BHV99}.  The collision nonlinearly triggers thermal instability in
the post-shock gas \citep{Hennebelle_Perault99} with the result that
the gas cools rapidly, producing a dense, cold, turbulent cloud
\citep{WF98, KI02, Heitsch_etal05, Heitsch_etal06, VS_etal06} which
soon becomes Jeans unstable \citep{VS_etal07, Heitsch_etal08}. Since
the interstellar medium, rather than homogeneous, is highly
structured, such shocked, cold, dense medium will naturally produce
clumps \citep{Bonnell_etal06}.  As the whole cloud collapses, the
column density increases rapidly, allowing the formation of molecular
gas \citep[e.g., ][]{HH08, Glover_etal10}.  From this point of view,
molecular clouds are in a global state of collapse, { with an internal
distribution of free-fall times due to the fluctuations in the density
field induced by the initial turbulence \citep{BH04, VS_etal06,
HH08}}. Note that 
this does {\em not} necessarily mean that the entire cloud is
collapsing along all dimensions; for example, in the toy model of the
Orion A cloud by \citet{HB07}, the angular momentum of the cloud
prevents collapse along the long dimension.

In this scenario, supersonic motions in molecular clouds are driven by
gravitational energy as the clouds proceed to collapse, a situation
that has been supported by \citet{HBH09}, who show that the synthetic
line profiles observed in numerical simulations of the process
resemble those in observations, e.g., the magnitude of the velocity
dispersion of \ala{13}CO line profiles, or the (small) core-to-core
velocity dispersion.

The key point, however, was emphasized by \citet{VS_etal07}, who
showed that the kinetic and gravitational energies of the collapsing
cloud develop a Virial-like relationship in complex
gravitationally-collapsing clouds, in which the absolute value of the
kinetic energy is very close to that of the gravitational energy
\citep[see Fig.~8 in][]{VS_etal07}.\footnote{{ Note that in
\citet{VS_etal07}, the absolute value of the gravitational energy is a
few times that of the kinetic energy. However, in that paper, the latter
energy was computed exclusively for the dense gas, while, for
simplicity, the gravitational energy was computed for the whole
numerical box.}}  In other words, the ``Virial'' relation between
kinetic and gravitational energy, rather than being an indicator of true
Virial {\it equilibrium}, simply shows the importance of gravity in
driving much of the non-thermal motions.

\citet{Field_etal08} have advanced a model of a ``gravitational
cascade'', which essentially captures the mechanism observed in the
simulations. Their model is analogous to a turbulent cascade, but in
which the quantity being cascaded is mass rather than kinetic energy,
and the main driver is gravity at all scales. In their model, these
authors propose that the contracting motions are somehow randomized, so
that the kinetic energy released by the collapse is converted into
quasi-isotropic motions, and Virial quasi-equilibrium can be established
at every scale (see their \S 3). Such Virial equilibrium states require
that the mass fragments are bound by an external pressure \citep{Elm89,
MT03, Field_etal10}.  In the latter paper, the authors write the Virial
equilibrium equation for a non-magnetized cloud of gas of mass $M$ and
radius $R$ in terms of its mass surface density $\Sigma = M/R^2$ as

\begin{equation}
  {\delta v^2\over R} = G\Sigma - {4\pi \over \Sigma} P_{\rm ext}
\label{eq:virial:sigma}
\end{equation}
where the first term in the right hand side corresponds to the
gravitational energy, and the second one is due to the external
pressure acting over the surface of the cloud.  However, the
successive virializations suggested by \citet{Field_etal08} will not
occur if cloud lifetimes are too short, as pointed out by
\citet{Bonnell_etal06}.

{ More importantly, a cloud undergoing collapse does not need to be
confined by an external pressure. Instead, the internal pressure
increases together with the density, and once the collapse has
advanced sufficiently,
the external pressure becomes negligible.  Indeed, as mentioned above,
these regions within molecular clouds are known to have much larger
thermal pressures than the ISM mean \citep[e.g., ][]{Blitz93}.  Thus,
rather than Virial equilibrium, the relevant principle here is that
the total energy of the system ($E_{\rm g}+ E_{\rm k}$) is
conserved. In this case, the ratio $dv^2/R$ is given by}

\begin{equation}
   {\delta v^2\over R} = {2 G\Sigma}.
\label{eq:collapse:sigma}
\end{equation}
Moreover, { if the collapsing scenario applies to all scales within
  molecular clouds, we expect eq.\ (\ref{eq:collapse:sigma}) to be
  valid not only for massive cores, but for molecular clouds in
  general.  Thus, it seems unavoidable to think that the molecular
  cloud supersonic linewidths, rather than being hydrodynamical
  turbulent motions, are what we refer to as {\it hierarchical
  gravitational collapse}, i.e, the local gravitational contractions
  occurring throughout the whole cloud, which, furthermore, is itself
  collapsing\footnote{Note that the resulting motions of such
  hierarchical gravitational collapse is, in a way, supersonic
  turbulence. However, rather than being the physical ingredient that
  opposes gravity, it is a consequence of gravity itself.}.  In other
  words, the kinetic energy gained during the hierarchical collapse
  must come from the gravitational energy released, and therefore
  develops a virial-like relation, except that the velocity dispersion
  is given by eq.  (\ref{eq:collapse:sigma}) rather than by the Virial
  relation $\delta v^2/R =G \Sigma$.}

\begin{figure}
\includegraphics[width=1\hsize]{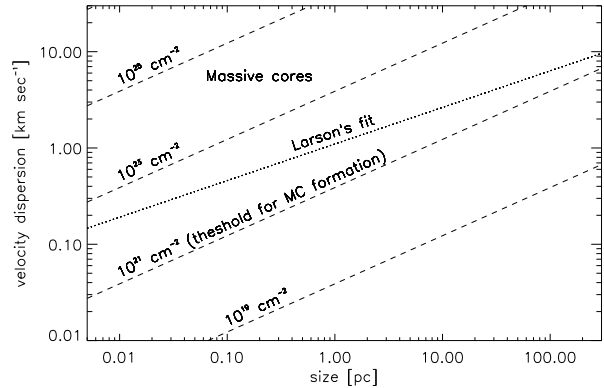}
\caption{Velocity dispersion-size space. The lines are the loci of
  cores and clumps at constant column density if the velocity
  dispersion is of gravitational origin, such that
  eq. (\ref{eq:collapse:sigma}) is valid.
\label{fig:dv_r_theory}}
\end{figure}

The assumption of gravity driving the chaotic motions in multiple
local centers of collapse, and thus developing a pseudo-Virial state
(eq. \ref{eq:collapse:sigma}), implies that massive, compact cores
should develop larger velocity dispersions for larger column densities
\citep[$N\sim$ \diezala{23}--\diezala{25}~cm\alamenos
2,][]{Shirley_etal03, Gibson_etal09, Wu_etal10}.  In
Fig.~\ref{fig:dv_r_theory} we show, in the velocity dispersion-size
space, lines of constant column density according to
eq.~(\ref{eq:collapse:sigma}), the locus of the \citet{Larson81}
relation, eq.\ (\ref{eq:larsonfit}), and the region where the massive
cores are located.  From this figure we note that typical local clouds
with mean column densities of \diezala{21}--\diezala{22}~cm\alamenos 2
will necessarily be close to Larson's relation, as observed in the
data \citep[e.g., ][]{Larson81, Heyer_etal09}.  Column densities far
below this relation will be unable to self-shield against background
UV radiation (e.g., \citet{HBB01}), and will be rapidly dissociated,
or do not form at all. However, as discussed previously, column
densities far above Larson's relation do exist - in the recently
massive compact cores, which occupy the locus $0.1 \le r/{\rm pc} \le
1$, $1\le \delta v/{\rm km\ s^{-1}} \le 10$ (see
Figs.~\ref{fig:dv_r_obs} and \ref{fig:dv_r_theory}) { and, although
they do not fall on Larson's relation, they do fall on Heyer's}.

\citet{Larson81} argued that supersonic hydrodynamics must be
important in cloud structure, and that the clouds cannot have formed
``by simple gravitational collapse''.  The model clouds by
\citet{VS_etal07} and \citet{HH08} do indeed exhibit important effects
of hydrodynamic turbulence, especially in early stages of formation.
However, hierarchical gravitational collapse comes to dominate at late
stages; it is just not ``simple'' because the geometry is complicated,
with several local centers of collapse within the parent cloud that is
contracting as a whole, giving rise to chaotic density and velocity
fields.  The local collapse occurs because the initial turbulence,
{ in combination with strong radiative losses} induces nonlinear
density fluctuations that have shorter free-fall times than the parent
cloud \citep{HH08}.

{ The scenario of hierarchical gravitational collapse is similar in
spirit to the classic notion of gravitational fragmentation
\citep{Hoyle53}, except that it takes place on a substrate populated
with nonlinear density fluctuations produced by the initial
turbulence. This eliminates the simultaneity of the collapses that would
be expected in a uniform medium. In this case, the driving force is
gravity, as opposed to previous scenarios where the confining force for
the (hydrostatic) clouds was the pressure external to the clouds
\citep[e.g.,][]{Elm89, MT03, Field_etal10}. It is interesting to note,
however, that in those works the presence of such a confining pressure
was {\it assumed}, and diverse mechanisms had to be devised in order to
attain the necessary high pressures, such as the weight of an atomic
component \citep{Elm89} or a recoil pressure from the photodissociation
regions around the clouds \citep{Field_etal10}. In our scenario, such
confining pressure does not exist, and the large internal pressure of
the clouds is simply a consequence of the ongoing collapse. }

\subsection{Avoiding the star formation rate conundrum}
\label{sec:avoid_ZP} 

The idea that GMCs and their substructure may be in a state of
gravitational collapse is not new. In fact, it was the first proposed
explanation for the observed supersonic linewidths
\citep{Goldreich-Kwan74}.  However, as it is well known, it was
quickly deemed untenable by \citet{Zuckerman-Palmer74}, who argued
that if all molecular gas in the Galaxy were in a state of free-fall,
the star formation rate (SFR) would be of order $M_{\rm mol}/t_{\rm
  ff}$, where $M_{\rm mol}$ is the total molecular gas mass in the
Galaxy, { and $t_{\rm ff}$ is the mean free-fall time of the molecular
gas}.  This results in a star formation rate roughly 100 times 
larger than the observed value of a few \Msun\ yr\alamenos1.

This simple reasoning, however, does not necessarily apply for real
molecular clouds, since it { neglects their complex, highly
  fragmented nature of their density distribution. Because of the
  existence of a wide distribution of local free-fall times in a GMC,}
the densest clumps collapse significantly sooner than the GMC at
large, and begin forming stars before the bulk of the GMC has
collapsed, as observed in simulations \citep[e.g.][]{VS_etal07,
  Heitsch_etal08, Banerjee_etal09, VS_etal10}. The feedback from the
stellar products (in the form of outflows, winds, ionizing radiation,
and SNe), can then interrupt the star formation (SF) process before it
exhausts the entire mass of the GMC. Furthermore, since it is well
known that most of the molecular gas mass is in the large-scale
structures, this implies that the SF efficiency will remain small,
because the stellar activity originates from the objects that collapse
first, which contain the minority of the mass. Indeed, numerical
simulations of GMC formation and evolution including self-consistent
stellar feedback by \citet{VS_etal10} show that the SF efficiency can
be maintained at realistic values throughout the evolution of the
cloud, still within the context of large-scale gravitational
contraction. As mentioned above, the youth and strong coevality of
stellar populations in the clouds, suggest that stellar feedback acts
quickly to suppress SF at any given location, and acts to either
relocate it, for a next generation of SF at a different location, or
even to completely disperse a cloud. Although the details of this
process still need to be refined, it is clear that the SF conundrum
does not need to apply for GMCs in general.


\section{Conclusions}\label{sec:conclusions}

In the present contribution we have discussed the nature of the
velocity dispersion-size relationship.  We showed that recent
observational results do not follow the standard \citet{Larson81}
velocity dispersion-size relation. Instead, (i) massive cores occupy
the locus defined by $0.1 \le R/{\rm pc} \le 1$; $1 \le \delta v/{\rm
km\ s^{-1}} \le 10$, and (ii) rather than a single $\delta v-R$
relation, the entire dataset for which independent mass estimates are
available seems to follow the \citet{Heyer_etal09} scaling, $\delta v
\propto \Sigma^{1/2}R^{1/2}$.  We showed that these results are
consistent with molecular clouds in a process of hierarchical
gravitational collapse, i.e., molecular clouds collapsing as a whole
while, at the same time, their cores collapsing locally, { creating
a complex supersonic velocity pattern that is however not fully
random, as in the standard notion of turbulence, but rather contains a
dominant globally contracting mode. As a consequence, this turbulence
cannot oppose the contraction, but rather feeds from it.}

We emphasized that, although hydrodynamic turbulence in the warm ISM
must play a role in producing the initial molecular cloud and its
condensations \citep{Clark_Bonnell05}, once gravity dominates motions
at late stages naturally results in the pseudo-Virial relationships
seen in the data.  Thus, it is not necessary (let alone likely) that
clouds and their cores are in pressure equilibrium with the external
medium, nor is it necessary to resort to some unspecified mechanism
that can accurately keep massive clouds with complex, non-spherical,
and clumped density distributions in an approximate equilibrium for
many crossing times.

\section*{Acknowledgements}
We thank to an anonymous referee for an encouraging and helpful
report.  This work has received partial support from grants UNAM/DGAPA
IN110409 to JBP, CONACYT 102488 to EVS, NSF AST-0807305 to LH and FH,
and CONACYT fellowship to MZ.  This work has made extensive use of the
NASA's Astrophysics Data System Abstract Service.

\label{lastpage}

 \end{document}